\documentclass[final,3p,times,twocolumn,authoryear]{elsarticle}

\usepackage{amssymb}
\usepackage{amsthm}
\usepackage{textcomp} 

\usepackage[]{hyperref}

\usepackage{lineno}

\journal{Earth and Planetary Science Letters}

\begin{document}

\begin{frontmatter}



\title{The global distribution of natural tritium in precipitation simulated with an Atmospheric General Circulation Model and comparison with observations}

\author[LMD,LSCE]{A.~Cauquoin\corref{cor1}}
\ead{Alexandre.Cauquoin@lmd.jussieu.fr}

\author[LSCE]{P.~Jean-Baptiste}
\author[LMD]{C.~Risi}
\author[LSCE]{{\'E}.~Fourr{\'e}}
\author[DAIS]{B.~Stenni}
\author[LSCE]{A.~Landais}

\address[LMD]{Laboratoire de M{\'e}t{\'e}orologie Dynamique (LMD), Universit{\'e} Paris 6, 4 place Jussieu, 75252 Paris Cedex 05, Tour 45--55, France}
\address[LSCE]{Laboratoire des Sciences du Climat et de l'Environnement (LSCE -- UMR 8212), Orme des Merisiers, 91191 Gif-sur-Yvette, France}
\address[DAIS]{Dipartimento di Scienze Ambientali, Informatica e Statistica (DAIS), Universit{\`a} Ca' Foscari Venezia, Via Torino 155, 30170 Venezia Mestre, Italy}

\cortext[cor1]{Corresponding author, phone number: +33144277387}

\begin{abstract}
The description of the hydrological cycle in Atmospheric General Circulation Models (GCMs) can be validated using water isotopes as tracers. Many GCMs now simulate the movement of the stable isotopes of water, but here we present the first GCM simulations modelling the content of natural tritium in water. These simulations were obtained using a version of the LMDZ General Circulation Model enhanced by water isotopes diagnostics, LMDZ-iso. To avoid tritium generated by nuclear bomb testing, the simulations have been evaluated against a compilation of published tritium datasets dating from before 1950, or measured recently. LMDZ-iso correctly captures the observed tritium enrichment in precipitation as oceanic air moves inland (the so-called continental effect) and the observed north-south variations due to the latitudinal dependency of the cosmogenic tritium production rate. The seasonal variability, linked to the stratospheric intrusions of air masses with higher tritium content into the troposphere, is correctly reproduced for Antarctica with a maximum in winter. LMDZ-iso reproduces the spring maximum of tritium over Europe, but underestimates it and produces a peak in winter that is not apparent in the data. This implementation of tritium in a GCM promises to provide a better constraint on: (1) the intrusions and transport of air masses from the stratosphere and (2) the dynamics of the modelled water cycle. The method complements the existing approach of using stable water isotopes.
\end{abstract}


\end{frontmatter}


\section{Introduction} \label{intro}

Model-data comparison of humidity fields and water fluxes is one method of judging the skill of Atmospheric General Circulation Models (GCMs) at modelling the hydrological cycle. However, many GCMs now also model the movement of the stable isotopes of water (H$_2$$^{16}$O, H$_2$$^{18}$O, HDO, H$_2$$^{17}$O), and these well-known tracers of the past and present-day hydrological cycle can also be used to validate model performance \citep{hoffmann98, mathieu, noone_simmonds, lee2007, schmidt, yoshimura, risi2010, werner2011}.

Tritiated water (HTO) is another useful tracer. Opposite to stable isotopes of water which show variations expressed in \textperthousand, driven by physical processes, HTO is rather a marker for reservoirs such as stratosphere, troposphere and oceans involved in the hydrological cycle. The content of tritium in these different compartments indeed varies by orders of magnitude. The concentration of tritium in precipitation is monitored by the Global Network of Isotopes in Precipitation (GNIP), and the database is available through \citeauthor{iaea}. A substantial dataset also exists of HTO concentration in polar ice (see \citet{fourre} and references therein). As a result, model-data comparison based on the HTO proxy is a promising way of increasing our understanding of GCM performance. 

Natural tritium is mainly produced by the interaction of galactic cosmic rays (GCR) with nitrogen of the upper atmosphere, at an average rate of $\sim$2500 atoms.m$^{-2}$.s$^{-1}$ \citep{craig_lal_1961, masarik09}. It has a radioactive half-life of 4500 $\pm$ 8 days \citep{lucas}. The vast majority of this natural (cosmogenic) tritium, which amounts to a global inventory of 3.6~kg, enters the hydrological cycle in the form of tritiated water molecules (HTO). This can be compared with the $\sim$520--550 kg of bomb tritium in the form of HTO \citep{michel, unscear} that was created in nuclear tests and subsequently caused a peak of bomb tritium in precipitation. Since the Nuclear Test Ban Treaty in 1963, the level of tritium measured in precipitation has been steadily decreasing due to radioactive decay and dilution in the world oceans. IAEA tritium long time series suggest that tritium levels in precipitation are close to their pre-nuclear test values.

It is estimated that 55\% of the natural production of tritium occurs in the stratosphere \citep{masarik99}. The natural-tritium content of stratospheric water vapour has been estimated to be some 5-9 x 10$^5$ TU (Tritium Units, where 1 TU corresponds to T/H ratio of 10$^{-18}$) \citep{ehhalt2002, fourre}, which is several orders of magnitude greater than the natural-tritium level in precipitation (a few TU only). This difference is due to the higher production rate and the lower water content in the stratosphere. This large stratospheric reservoir also makes tritium an extremely valuable tracer for mapping the intrusion of stratospheric air masses into the troposphere, in particular in Antarctica, a region under the influence of the polar vortex \citep{taylor, jouzel1979, jouzel1982, wagenbach, fourre}.

\citet{koster} pioneered the implementation of tritium in a GCM. In the only study published so far, these authors simulated the bomb-produced tritium atmospheric distribution with the GISS GCM. They used a simple bomb-tritium input function, a low model resolution ($8^{\circ} \times 10^{\circ}$ and 9 vertical layers) and only a very short simulation period (30 July days). They did not study the latitudinal variations of tritium concentration in the tropospheric vapour and precipitation. In addition, the low model resolution and short simulation period prevented the detailed study of continental recycling. Further, the correct representation of stratospheric air inputs in a GCM requires a higher vertical resolution than could be used by \citeauthor{koster}, especially above the tropopause. Simulations must also be longer than one year to study of its seasonal cycle. Finally, tritium data for the \citeauthor{koster} study were much more scarce than now, which restricted the model evaluation. With a current state-of-the-art GCM run on a modern computer, these limitations need no longer apply. 

This paper describes the implementation of tritium in the Laboratoire de M{\'e}t{\'e}orologie Dynamique Zoom (LMDZ) Atmospheric General Circulation Model developed at LMD \citep{hourdin2006, hourdin2013}. Our objective is to test the simulated distribution of HTO in the atmosphere due to climate processes; we therefore focus on natural tritium and run the simulations without the massive increase of tritium due to nuclear weapon testing. This restriction allows us to study the transport of tritium and its spatial/seasonal variability under steady-state tritium inputs.

Here, the simulation of the natural-tritium content of water vapour and precipitation is validated against a compilation of published tritium datasets of pre-bomb samples dated before 1950 and of recent samples for which the bomb component can be neglected.

The outline of this paper is as follow. In Section \ref{model_description}, we give a description of the isotopic processes included in the model. We also describe how the model has been adapted for tritium implementation, the various simulations performed and the tritium datasets used for validation. In Section \ref{validation}, we present simulated spatial variations of HTO in precipitation and water vapour and validate them against tritium data. We also evaluate the seasonal variability linked to exchanges of air masses between the stratosphere and the troposphere over two well-documented areas, Europe and Antarctica. In Section \ref{discussion}, we use the patterns of observed and modelled spatial variability of HTO to identify its driving mechanisms for present-day climatology and apply specific sensitivity tests on these two regions.

\section{Model simulations and datasets} \label{model_description}

\subsection{LMDZ-iso and isotopic processes}
\citet{risi2010} have already implemented water isotopes in the LMDZ model in the version LMDZ4 \citep{hourdin2006}. Since then, developments improved the capabilities of the model and we have implemented tritium in version LMDZ5a \citep{dufresne}, whose isotopic version is hereafter referred to as LMDZ-iso. The physical parameterizations of LMDZ5a are very similar to those used in LMDZ4, with the same Emanuel convective parameterization \citep{emanuel} coupled to the \citet{bony_emanuel} cloud scheme. The advection of water in its vapour and condensed states is calculated using the \citet{van_leer} scheme. The dynamical part is based on a discrete latitude-longitude grid, at the standard resolution of $2.5^{\circ} \times 3.75^{\circ}$. To ensure a realistic description of the stratosphere and of the Brewer-Dobson circulation, the model is run with 39 layers in the vertical (against 19 in the LMDZ4 version), including 22 located above the 200~hPa pressure level \citep{lott}. 
 
Following \citet{risi2010}, LMDZ-iso uses the same physical and dynamical descriptions of the HTO molecule as for the other water isotopes. Due to mass and symmetry differences, the various isotopic forms of the water molecule (H$_2$$^{16}$O, H$_2$$^{18}$O, HDO, H$_2$$^{17}$O, HTO) have slightly different physical properties (e.g., saturation vapour pressure, molecular diffusivity), and are therefore redistributed between the vapour and condensed phases at each phase change differently, depending on atmospheric conditions (temperature, vapour saturation, etc). The implementation of the water stable isotopes H$_2$$^{16}$O, H$_2$$^{18}$O and HDO in LMDZ-iso and of their associated isotopic processes (kinetic and equilibrium) have been extensively described by \citet{risi2010} and \citet{bony2008}. Contrary to stable isotopes of water, the uncertainties of these isotopic effects are usually considered of minor importance because of the larger differences in tritium content in water between the reservoirs.

The saturated vapour pressure of heavier molecules (e.g., HTO) differs from that of lighter ones (e.g., H$_2$$^{16}$O) creating equilibrium fractionation during a change of phase. The temperature-dependent coefficients for equilibrium fractionation $\mathrm{\alpha_{eq}}$ between vapour and liquid water or ice for HTO are given by \citet{koster}. Equilibrium fractionation has a significant effect on the variations of the ratio HTO/H$_2$$^{16}$O in the precipitation. As an example for a temperature of 10$^{\circ}$C, the coefficient for liquid/vapour fractionation, $\mathrm{\alpha_{eq}}$, is equal to 1.2378 for HTO/H$_2$$^{16}$O fractionation, compared to $\mathrm{\alpha_{eq}}$ values of 1.0107 and 1.098 for H$_2$$^{18}$O/H$_2$$^{16}$O and HDO/H$_2$$^{16}$O fractionations, respectively.

Isotopic fractionation is also introduced during evaporation of water at the ocean surface (with the ocean considered as an infinite reservoir). As for the water stable isotopes, the isotopic evaporation of HTO is computed following \citet{merlivat_jouzel_1979}, including equilibrium fractionation and kinetic effects depending on the surface wind speed.

Re-evaporation and diffusive exchanges between rain and vapour as the rain falls through the atmosphere are taken into account in the model. At any level in the vertical, the isotopic composition of the precipitating water depends on several parameters: (1) the composition of the condensed water that has been converted to precipitation at upper levels, (2) the fraction of precipitation that has re-evaporated and (3) the diffusive exchanges that take place between the falling drops and the downdraft water vapour. The re-evaporation and diffusion processes are calculated depending on relative humidity following \citet{bony2008}, who extended the work of \citet{stewart}. The calculation includes the kinetic effects associated with the re-evaporation of the drops in an unsaturated environment. If the relative humidity is 100\%, we simply assume total re-equilibration between raindrops and the surrounding vapour \citep{risi2010}. The fractionation processes linked to re-evaporation of precipitation are extensively described by \citet{bony2008}.

In addition to equilibrium effects when vapour condenses in the frozen form (at temperature, $T$, below $-15^{\circ}$C), the air is supersaturated with respect to ice. Effective fractionation is then expressed as a function of the equilibrium fractionation factor between vapour and ice, molecular diffusivities, and the supersaturation over ice, $S$. In LMDZ-iso, $S = 1 - 0.004T$ \citep{risi2010}. The effective fractionation $\mathrm{\alpha_{eff}}$ is taken from \citet{jouzel_merlivat}:

\begin{equation}
     \mathrm{\alpha_{eff}} = \mathrm{\alpha_{eq}} \left[\frac{S}{1 + \mathrm{\alpha_{eq}} (S-1) D/D'} \right]
\end{equation}
where $D$ and $D'$ are the molecular diffusivities in air of H$_2$$^{16}$O and of the heavier isotope. Here, the diffusivity of HTO is set to 0.968 times that of H$_2$$^{16}$O, derived from \citet{merlivat1978}.

\subsection{Implementation of tritium production} \label{3H_prod}
\begin{figure}[htb]
\center
\includegraphics[width=0.95\columnwidth]{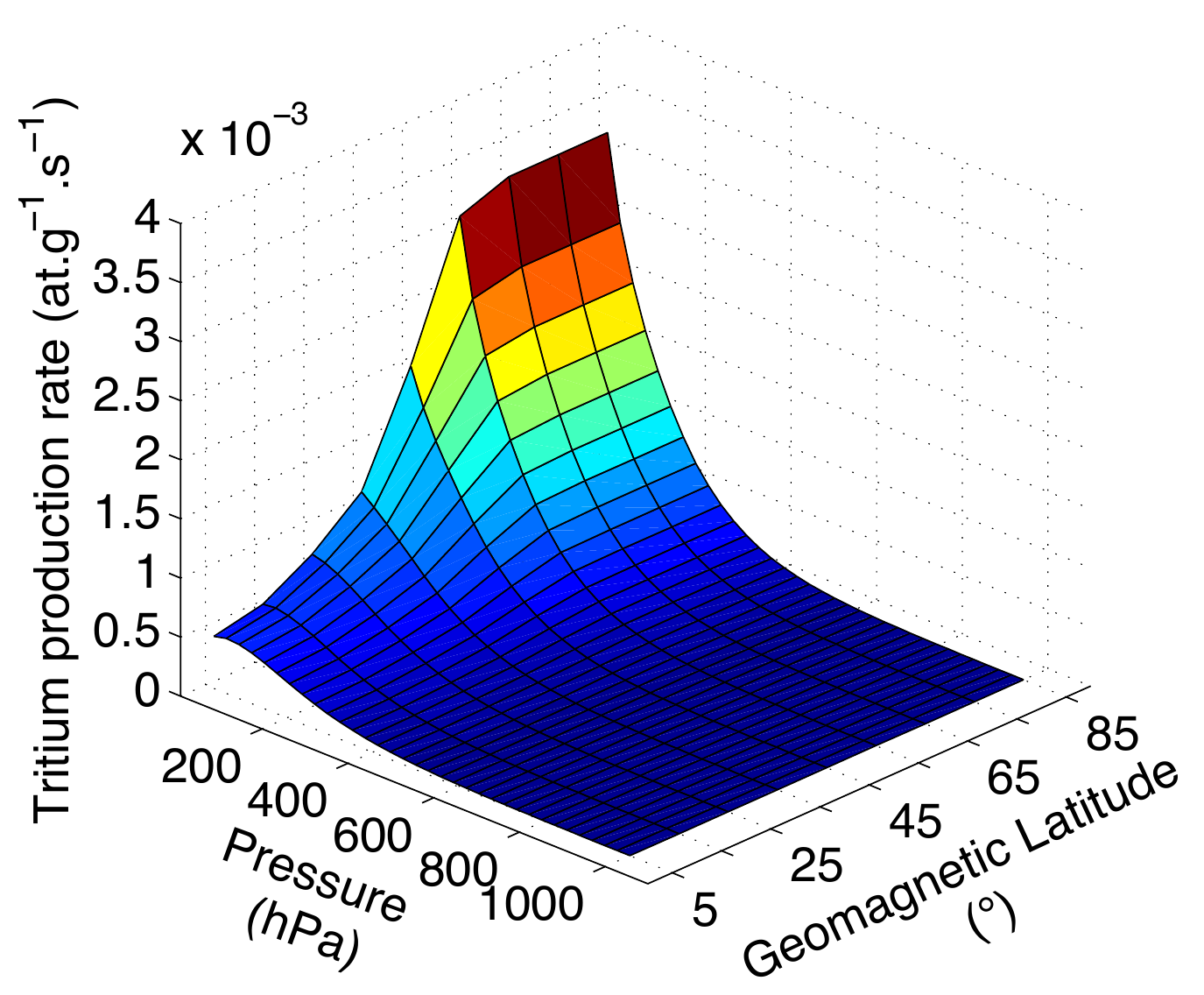} 
\caption{Natural tritium production rate (colour scale in atoms.g$^{-1}$.s$^{-1}$) according to \citet{masarik09} for a solar modulation parameter $\phi$ = 550 MV and present-day geomagnetic field intensity.}
\label{fig1}
\end{figure}

The production of natural cosmogenic tritium is induced by the nuclear interactions between the Galactic Cosmic Rays (GCR) (high-energy charged particles and their secondary products) and atmospheric nitrogen atoms. The higher the solar or geomagnetic field intensity, the more GCR particles are deflected, which leads to a decrease of cosmogenic production. Therefore, the production of cosmogenic isotopes is inversely related to solar activity (heliomagnetic field) and to the Earth's magnetic field intensity. The production of natural tritium has a strong latitudinal dependency caused by geomagnetic modulation, with a stronger shielding effect at the geomagnetic equator \citep{lal_peters, masarik09}. The dependency with altitude is expressed by the trade-off between the energy of the secondary particles produced by GCR (decreasing with the atmospheric pressure due to spallation reactions and geomagnetic rigidity) and their numbers (increasing with the atmospheric pressure). The production rates of the natural tritium (Fig.\ \ref{fig1}) are taken here from the revised calculations of \citet{masarik09} for a long-term average value of solar activity variations (described as the solar modulation potential e.g., loss of average rigidity of a particle in the heliosphere) $\phi$ = 550~MV \citep{castagnoli} and a relative geomagnetic field intensity $B$ = 1 (corresponding to the present-day value). These productions rates have been deduced by combining calculated neutron and proton spectra with experimental or evaluated cross sections of the cosmogenic isotopes. Here, we assume that one atom of tritium produced corresponds to one molecule of HTO created. 

\subsection{Simulations and sensitivity tests} \label{simulations}
Contrary to \citet{risi2010} who used the three-dimensional horizontal winds from ERA-40 reanalyses \citep{uppala_ERA40}, all our simulations were nudged by the horizontal winds from 20CR reanalyses \citep{compo_20CR}. Whereas the ERA-40 reanalyses go back only to 1958, the 20CR reanalyses go back to 1871; in due course, the latter dataset will thus allow us to investigate the bomb tritium peak of the 1950s-60s. Prior to the first run with tritium in LMDZ-iso, the model was spun-up for the period between 1942 and 1989 --- more than three times the tritium half-life. Comparison between the mean globally averaged tritium concentration in precipitation in 1989 (end of the spin-up) and over 1990-2008 indicates that this spin-up time is sufficient to allow tritium to reach steady state. The model was evaluated using the outputs for the period 1990-2008. We averaged the simulations over this full period to produce the annual-mean spatial distributions.

We performed several simulations with different input parameters in order to test the sensitivity of the model to tritium production rate, tritium concentration at the ocean surface and the description of fractionation processes. \citet{masarik99, masarik09}, hereafter referred to as MB, evaluated the average error on calculated production rates to be within 30-50\%. To test the consequences of such an uncertainty on the estimated tritium in precipitation in LMDZ-iso, we performed simulations with tritium production rates equal to -30\%, 0\% and +30\% of the nominal production rate. The ocean being an important sink of HTO, we performed simulations with uniform tritium contents at the ocean surface of 1~TU and 0.2~TU, representing an upper and lower boundary of the estimated natural value, respectively (see Table S1 in supplementary material). These sensitivity tests allowed us to tune the model parameters as closely as possible to the available observations. Finally, we also evaluated the influence of fractionation processes on natural HTO content by performing a simulation without isotopic fractionation. 

To summarize, the main simulations were as follows:
\begin{enumerate}[(1)]
\item ``1TU\_MB'': with the MB natural tritium production rate function (see Fig.\ \ref{fig1}) and a tritium concentration at the ocean surface set to 1 TU \citep{begemann}; 
\item ``0.2TU\_MB'': as (1) but with the ocean surface set at 0.2 TU, corresponding to an extrapolated estimation of pre-bomb tritium in the North Atlantic surface waters \citep{dreisigacker}; 
\item ``1TU\_MB+30\%'': as (1) but with the MB tritium production rate increased by 30\%; 
\item ``1TU\_MB-30\%'': as (1) but with the MB tritium production rate decreased by 30\%;
\item ``0.2TU\_MB-30\%'': with the ocean surface set at 0.2 TU and the MB tritium production rate decreased by 30\%;
\item ``$\alpha$ = 1'': same as (5) but without isotopic fractionation (equilibrium + kinetic).
\end{enumerate}

\subsection{Description of the available tritium data} \label{data}
To evaluate the spatial distribution of our modelled cosmogenic tritium, we need a wide spatial coverage of tritium data for periods when the bomb component is low enough to neglect. To evaluate the seasonal variations, well-resolved seasonal cycles are also needed. We present here the different tritium datasets that were used. For the distribution of natural tritium in precipitation, we firstly exploit available pre-bomb data in rainwater and fresh water reservoirs, as well as some pre-bomb tritium data retrieved from ice cores (referenced in Table S1). Also included in this dataset are a few stations from the GNIP database. These data allow the accurate extrapolation of the tritium temporal trend to the subsequent decades (also referenced in Table S1). Although this measurement network has been run down, the several decade-long runs of data allow the decreasing bomb-tritium signal to be extrapolated to obtain the asymptotic level, i.e., the natural level, even if low level tritium measurements have ceased. However, because of their limited number, we have also selected, when necessary, recent data for tritium in precipitation from the \citeauthor{iaea} database. To make sure that the bomb-component in these profiles is vanishing, we checked that the difference in tritium content between 2000 and 2008 (date of last available tritium measurements in IAEA stations) was negligible. for which the bomb-tritium component is vanishing. For the evaluation of the annual-mean spatial distribution of tritium simulated by LMDZ-iso, we have selected the year 2002. This year is a good compromise between the quantity of available measurements and the influence of bomb tritium being as small as possible. In the same way, we restricted the GNIP data to the years after 2000 to study the seasonal variations of tritium. Focusing on the tritium distribution over the Antarctica, we therefore also considered tritium data of surface snow samples (1 m depth integrated sample) from two ITASE (International Trans-Antarctic Scientific Expedition) traverses \citep{proposito, becagli}, carried out during the Southern Hemisphere summers of 1998/99 and 2001/02. Tritium concentration was measured by liquid scintillation counting \citep{proposito} and the associated errors range from $\pm$1~TU (2$\sigma$) at background levels, to $\pm$3~TU at the level of 35~TU. These traverses connected well-known East-Antarctic sites such as Terre Ad{\'e}lie, Terra Nova Bay, Dome C and Talos Dome.

\section{Results of the model-data comparison} \label{validation}

\subsection{Vertical profile} \label{vertical}
\begin{figure}[htb]
\center
\includegraphics[width=0.95\columnwidth]{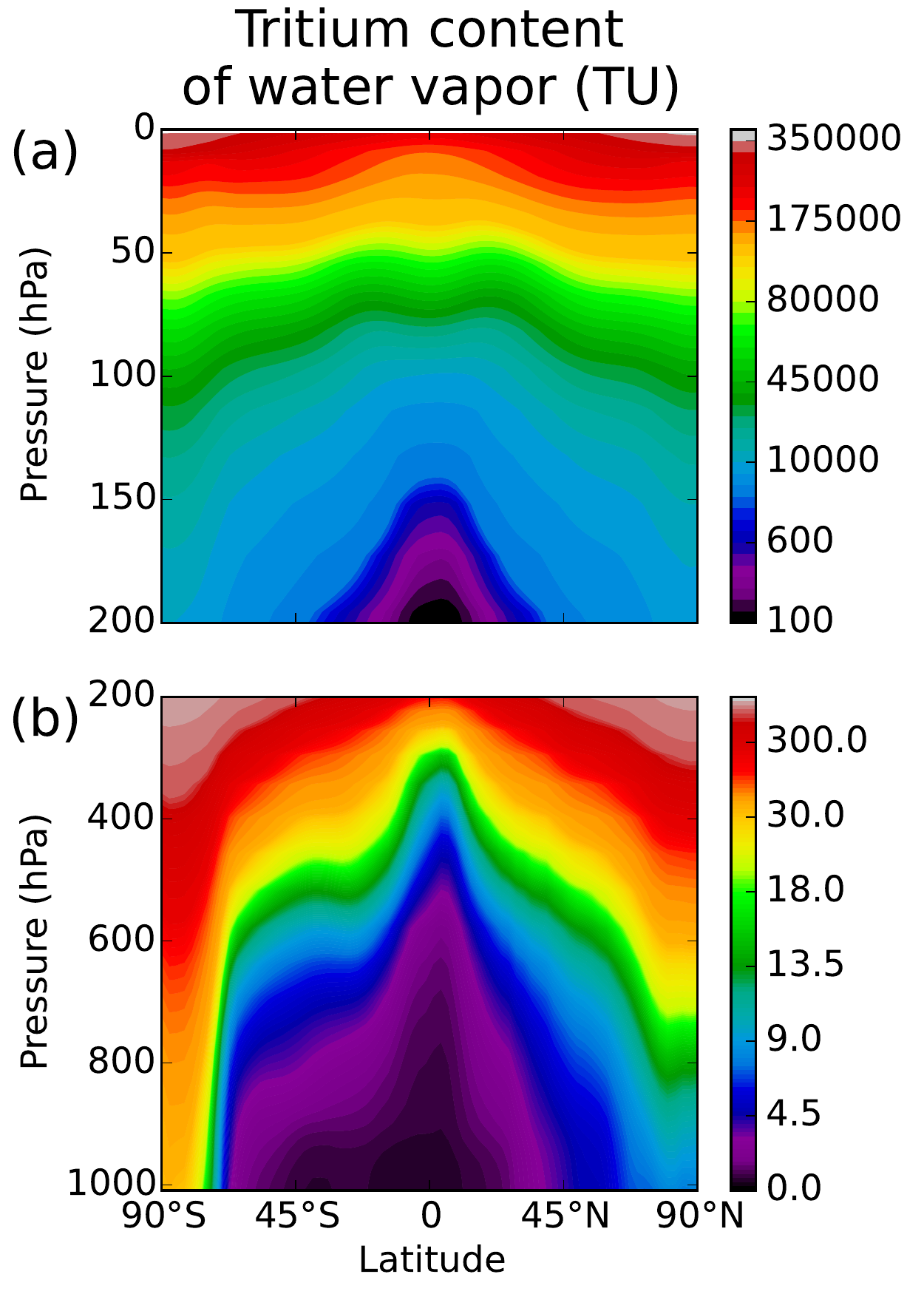} 
\caption{Annual-mean tritium content of water vapour as a function of the atmospheric pressure (hPa) and latitude in LMDZ-iso (``0.2TU\_MB-30\%'' simulation): (a) the stratosphere between the top layer and 200~hPa (b) and the troposphere between 200 and 1000~hPa.}
\label{fig2}
\end{figure}

Fig.\ \ref{fig2} shows the annual-mean spatial distribution of the modelled tritium content of water vapour from the ``0.2TU\_MB-30\%'' simulation as a function of atmospheric pressure and latitude. This simulation with the lowest input is the closest to the data. As expected, while the tritium level in water vapour in the first vertical layers of the model above the ground is a few TU only (Fig.\ \ref{fig2}b), tritium levels in the stratosphere are higher by several orders of magnitude (10$^5$~TU, Fig.\ \ref{fig2}a). The average modelled tritium content in the stratosphere (defined as above the 200~hPa pressure level) varies between 1.12 and 2.08 $\times$ 10$^5$~TU according to the input parameters used in LMDZ-iso (see Section \ref{simulations}). These high values are due to the higher tritium production rate (Fig.~\ref{fig1}) and to the dryer air at higher altitude. One can also notice the strong latitudinal dependency of tritium levels in the stratosphere, with an increasing trend from the Equator to the poles. This is expected because the cosmic ray flux (and therefore the natural-tritium production rate) is stronger at high geomagnetic latitudes. Model values are on the low side of the range of values estimated by \citet{ehhalt2002} and \citet{fourre}, 5.10$^5$~TU and 9.10$^5$~TU respectively, based on a small number of stratospheric measurements made prior to 1984. This stratospheric tritium dataset is very limited in time (seasonality) and space (latitude) and the decreasing trend of stratospheric bomb tritium must be extrapolated over several decades and several orders of magnitude to estimate the natural value. These factors make the uncertainty in these extrapolated values difficult to assess; yet they may be quite high. Clearly, new stratospheric measurements would be necessary to see whether or not the model really underestimates the stratospheric production.
\begin{figure*}[htb]
\center
\includegraphics[width=0.95\textwidth]{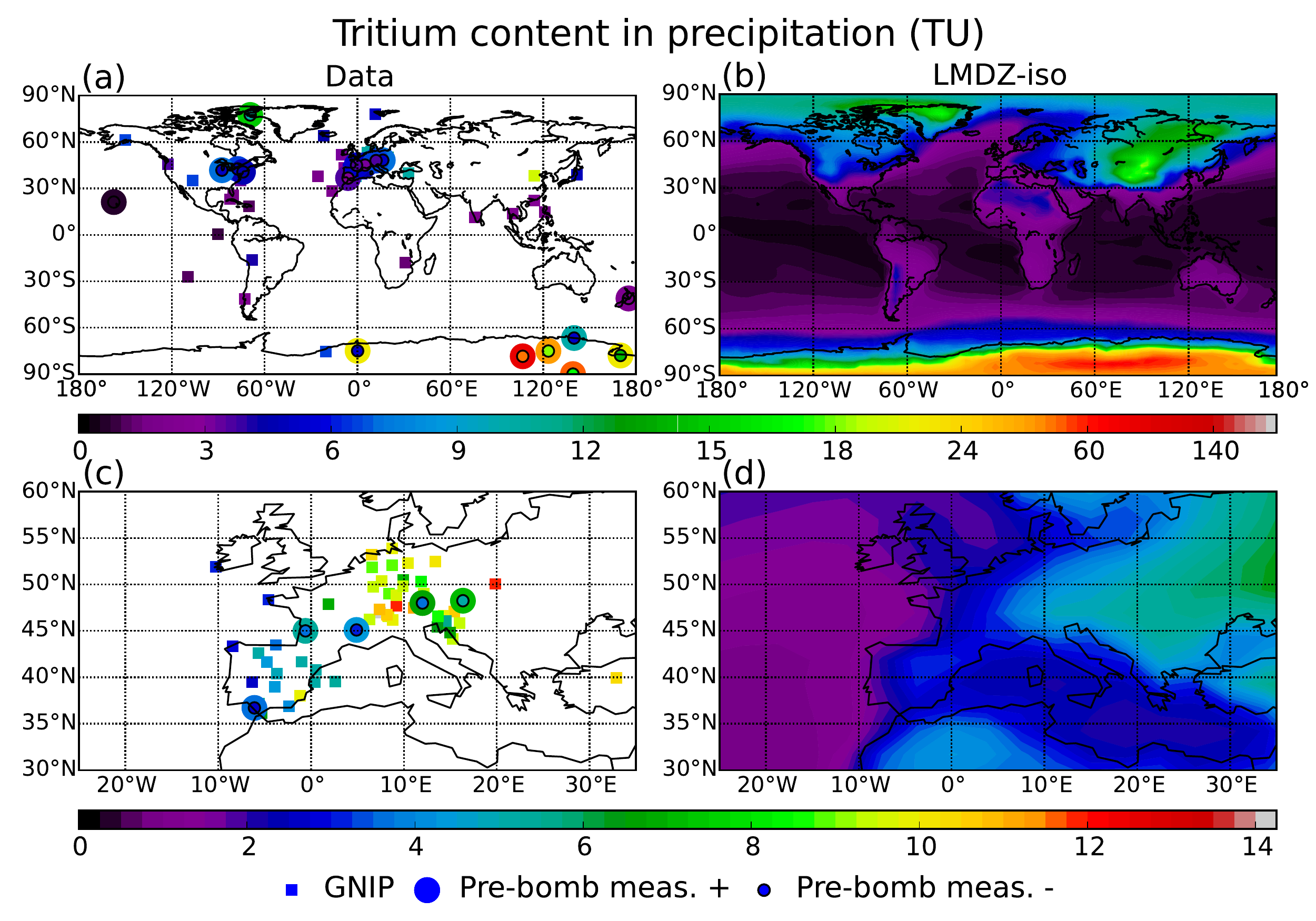} 
\caption{(a) Annual-mean tritium in precipitation from the data and (b) from LMDZ-iso ``0.2TU\_MB-30\%'' simulation. Note that the colour scale is not linear. (c) and (d) as (a) and (b) but for the European region, where the GNIP data are the most abundant. The data used here are the GNIP data (squares) and the pre-bomb measurements reported in Table S1 (large and small circles for the higher and lower limits respectively).}
\label{fig3}
\end{figure*}

\subsection{Spatial distribution of tritium in precipitation} \label{hto_world}
Fig.\ \ref{fig3}a and \ref{fig3}b compare the global spatial distribution of our modelled tritium in precipitation from the ``0.2TU\_MB-30\%'' simulation with the GNIP data and the pre-bomb measurements. The figures show that the spatial distribution of the annual-mean tritium in precipitation is well simulated by the model. In particular, both data and model show the clear contrast between the marine and the continental stations, with lower values being found over the oceans. Classically, this continental effect is explained by the production of tritium incorporated in precipitation while tritium in marine re-evaporation remains low due to the buffering effect of the ocean. It is well simulated, with an increasing trend from the European coasts (2-3~TU) to Asia (10-20~TU) for example (Fig.\ \ref{fig3}c, d) (see details in Sections \ref{sensitivity} to \ref{europe}). Antarctica also shows an increasing trend moving inland from the coast, especially on the East Plateau. The results for this area are analyzed in detail in Section \ref{antarctica}. In contrast, but in agreement with the data, Greenland does not show such high tritium concentrations, because of the greater influence of marine air masses.

\subsection{Seasonal variability of tritium in precipitation} \label{seasonal}
We compare here the modeled and measured seasonal variability of HTO in precipitation at two sites representative of each hemisphere : Vienna (Europe) and Halley Bay (Antarctica).

\begin{figure}[htb]
\center
\includegraphics[width=0.9\columnwidth]{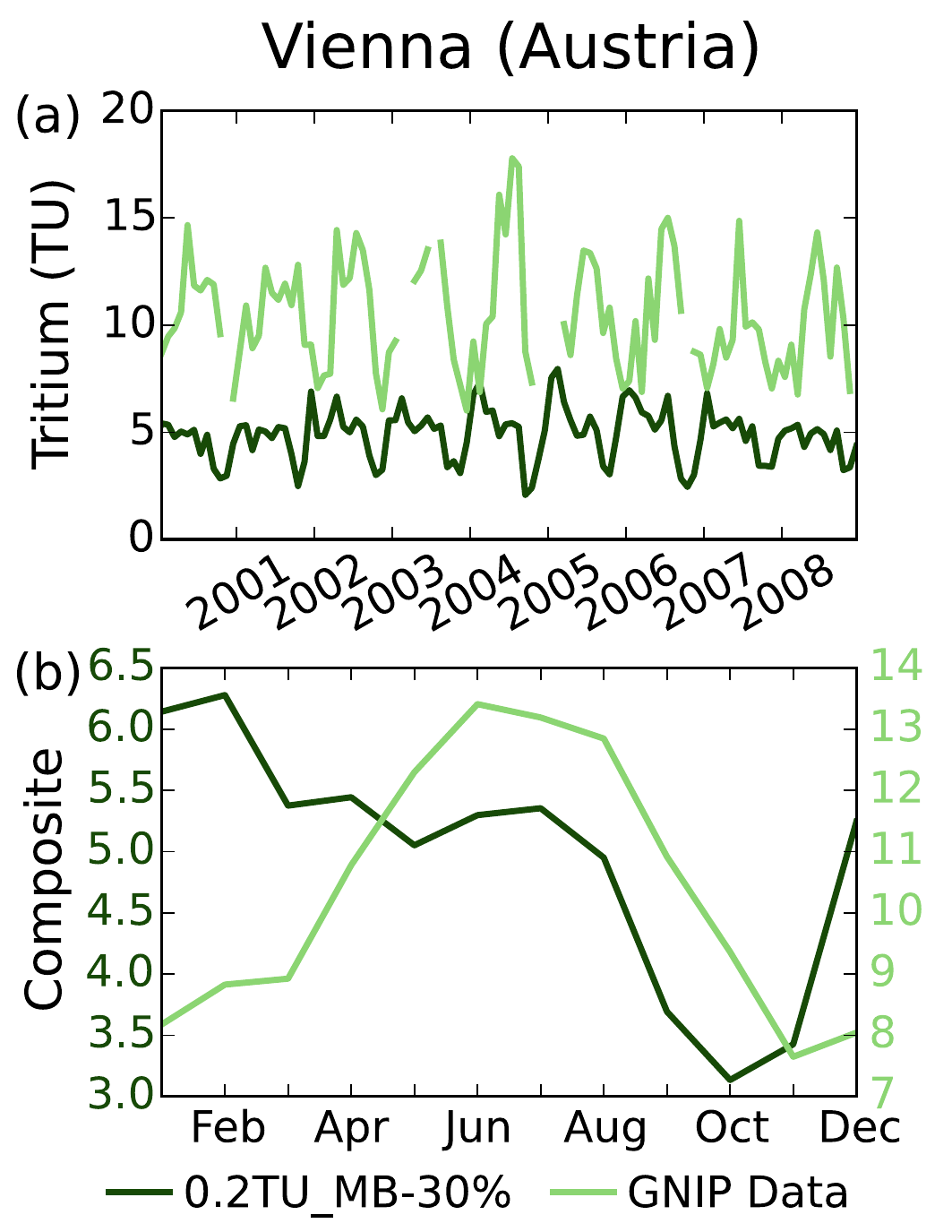} 
\caption{(a) Tritium in precipitation between 2000 and 2008 and (b) the corresponding monthly composites at the Vienna station. The light and dark green curves correspond to the GNIP data and the ``0.2TU\_MB-30\%'' simulation respectively. Beware that the simulated and observed signals are not plotted on the same scale.}
\label{fig4}
\end{figure}

One of the best-documented GNIP stations is Vienna (Austria, 48$^{\circ}$ 12' N, 16$^{\circ}$ 22' E). The seasonal variability of tritium in precipitation measured at this station between 2000 and 2008 (Fig.\ \ref{fig8}a and the corresponding monthly composite in Fig.\ \ref{fig4}b) shows a clear seasonal cycle with a spring-summer maximum (between April and September, Fig.\ \ref{fig4}b). The same tritium seasonal cycle is observed in the other European stations of the GNIP network. This maximum, referred to as ``Spring-Leak'', is classically explained by the exchange of tropospheric and stratospheric air masses occurring predominantly during late winter and spring in the region of baroclinic zones and tropopause discontinuities of the mid-latitudes \citep{newell}. Both model and data show similar relative amplitudes, however the simulated values of tritium in precipitation at this station are lower probably because some bomb-tritium is still present in the data. Still, the model is able to reproduce relatively high values of tritium in precipitation during spring but the highest values occur in winter. We do not have a definitive explanation for the mismatch between model and data but the stratosphere-troposphere exchange is clearly a complex process that appears to be neither geographically nor seasonally uniform \citep{jordan}. Actually, studies performed with a GCM \citep{pedro2011} or particle dispersion model \citep{james2003, stohl2006} have found that the modelled maximum intrusion of stratospheric air in the troposphere in the Arctic and Northern Hemisphere occurs in winter as simulated in our model. The more frequent occurrence of rapidly descending air in winter is explained in these aforementioned studies by the intensity of baroclinic systems being greatest in wintertime \citep{james2003}. This process is probably too strong in our model, which would explain the model-data mismatch observed for the seasonal cycle in Europe. Further model-data comparisons are thus needed to refine the exact mechanism dominating the intrusion of stratospheric water at the seasonal scale in the model and in the real world. 

\begin{figure}[htb]
\center
\includegraphics[width=0.9\columnwidth]{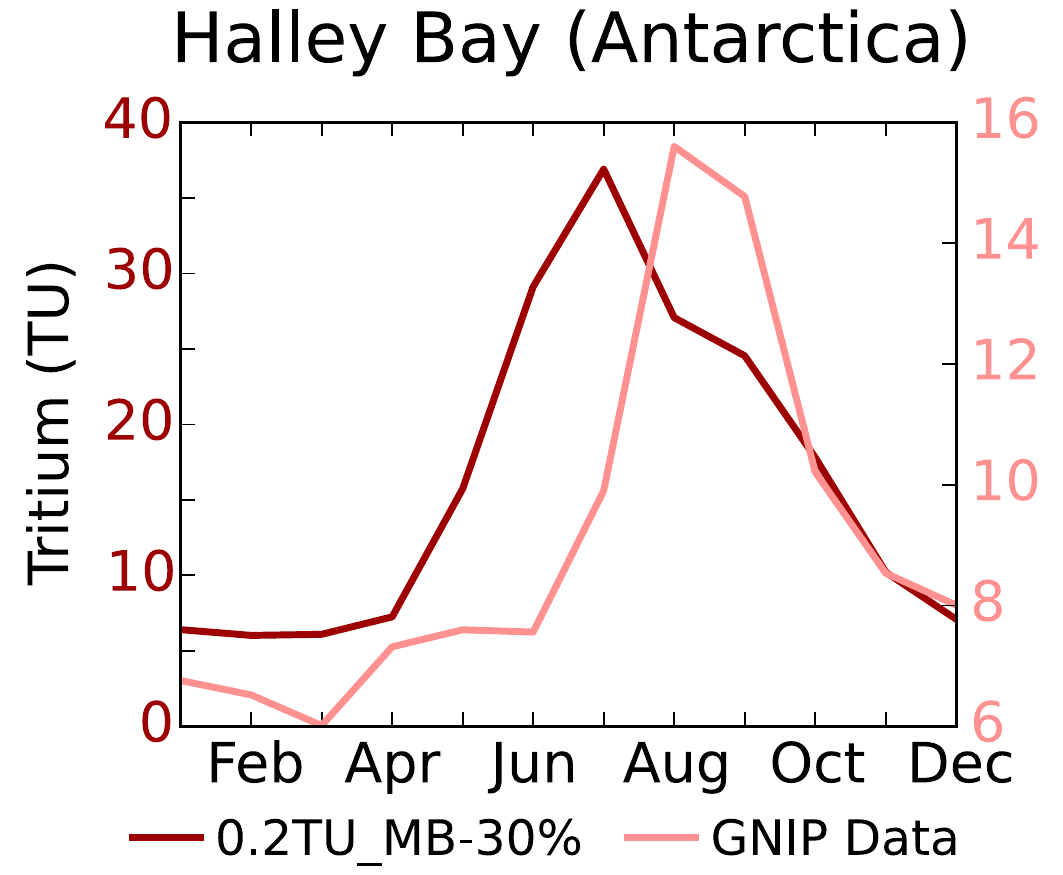} 
\caption{Monthly composite of measured tritium-in-preciptation (light-red) and of simulated tritium in precipitation (dark-red) at Halley Bay (Antarctica). Beware that the simulated and observed signals are not plotted on the same scale.}
\label{fig5}
\end{figure}

At the coastal Antarctic station of Halley Bay (75$^{\circ}$ 35' S, 20$^{\circ}$ 34' W), Fig.\ \ref{fig5} shows that the seasonal maximum in tritium fallout modelled by LMDZ-iso takes place during the austral winter (April to September) when the air is particularly dry and the Antarctic inland is isolated from marine air masses by the polar vortex. This result is also fully consistent with ice-core observations of the peak of pre-bomb tritium occurring in September \citep{jouzel1982}. This confirms that, in contrast to the situation in the Northern Hemisphere, the model correctly simulates the strongest input of stratospheric moisture during winter due to a strong descent through the polar vortex and possibly through the sedimentation of polar stratospheric clouds  \citep{wagenbach}. Still, the seasonal amplitude of the tritium in precipitation or in ice as calculated by the model is almost three times higher than in the observations over this coastal station. One explanation for this effect could be that the tritium-rich intruding air in LDMZ-iso spreads out too much in the horizontal \citep{hourdin1999}. It thus reaches the coastal Antarctic area and the near ocean that should not otherwise be much influenced by these stratospheric intrusions.

\section{Discussion: what drives the spatial variability of tritium in precipitation?} \label{discussion}

In this section, we use the patterns of observed and modelled spatial variability of HTO to identify its driving mechanisms. We report the influences of tritium production rate, tritium concentration at the ocean surface and fractionation processes on the global distribution of tritium in precipitation, and especially how they affect the so-called continental and latitudinal effects occurring in the model. We also apply these sensitivity tests on two well-documented areas, Europe and Antarctica.

\subsection{Influences of tritium production and concentration at the ocean surface on the global distribution of tritium in precipitation} \label{sensitivity}
\begin{figure}[htb]
\center
\includegraphics[width=0.95\columnwidth]{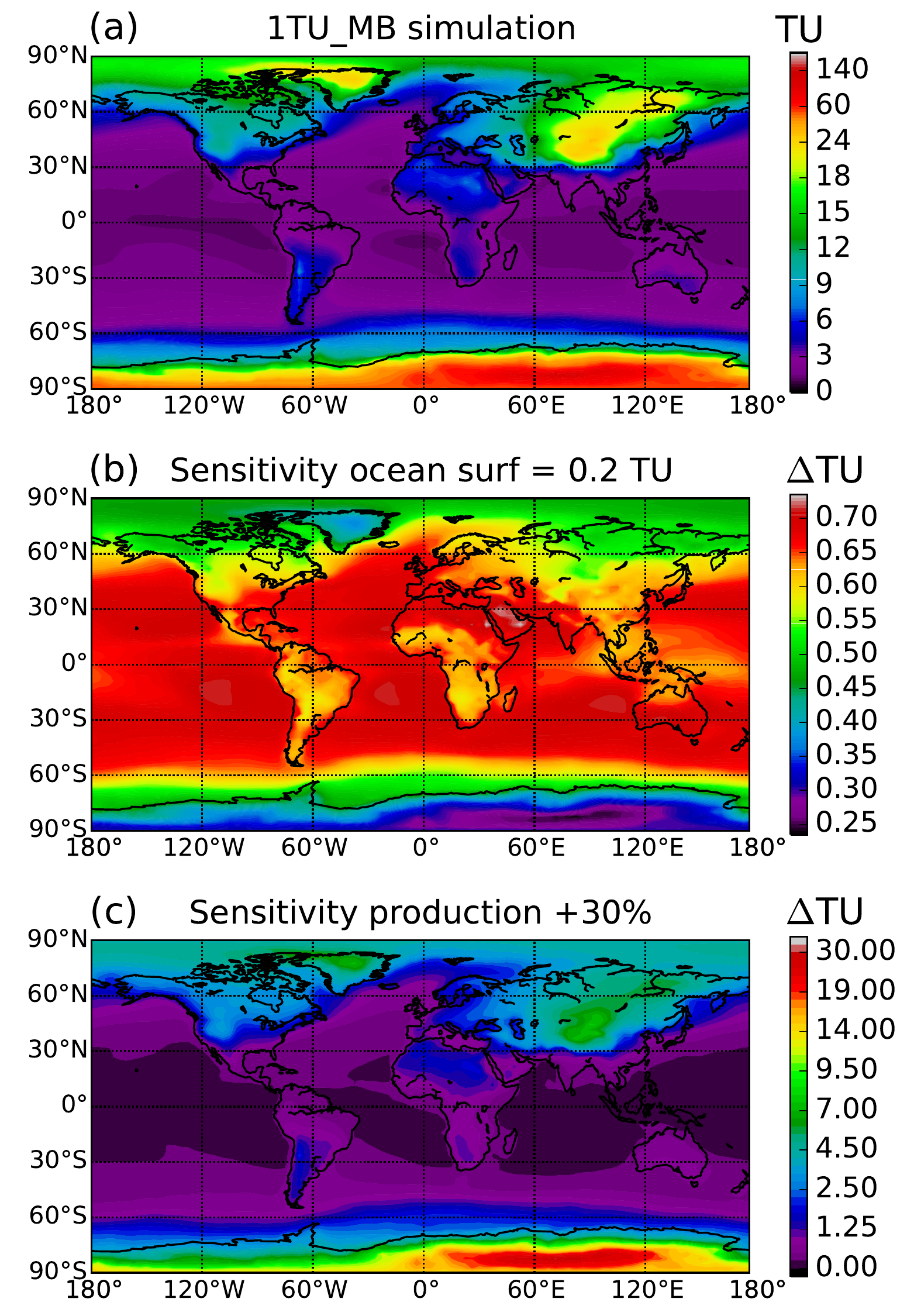} 
\caption{(a) Annual-mean tritium in precipitation in the ``1TU\_MB'' simulation (control). (b) Results from the ``1TU\_MB'' simulation minus results from the ``0.2TU\_MB'' simulation. (c) Results from the ``1TU\_MB+30\%' simulation minus results from the ``1TU\_MB'' simulation.}
\label{fig6}
\end{figure}

In order to identify the processes linking the spatial variability of tritium in precipitation to the production rate and the tritium concentration at the ocean surface, we report in Fig.\ \ref{fig6} the differences between the global distribution of tritium from the ``1TU\_MB'' simulation (Fig.\ \ref{fig6}a) and the ones from the ``0.2TU\_MB'' and ``1TU\_MB+30\%'' simulations (see Section \ref{simulations} for a description of the simulations). The global content of tritium is, as expected, lower in the ``0.2TU\_MB'' simulation than in the control one, ``1TU\_MB'' (Fig.\ \ref{fig6}b). The difference between the two simulations mainly occurs over the oceans. Over the continents, the difference in tritium content between the ``1TU\_MB'' and ``0.2TU\_MB'' simulations decreases as the distance from the coast increases. This suggests that in continental regions, the source of tritium in precipitation is mostly the tritium production in the high atmosphere with only relatively little contributed by oceanic evaporation.

The same contrast between the relative source-strength of the oceanic tritium background level and the tritium production in the high atmosphere is observed when looking at the difference in tritium distribution between the ``1TU\_MB+30\%'' simulation and the ``1TU\_MB'' simulation in Fig.\ \ref{fig6}c. Over oceans and coastal regions, there are no significant differences between the two simulations because of the stronger influence of the oceanic surface water on the tritium signal. On the contrary, the difference of tritium level between the two simulations increases by 3--8~TU at mid-latitude in continental regions confirming that this continental effect of tritium increase is mostly due to the cosmogenic tritium production term. In addition, we observe that the difference between the two simulations increases at higher latitudes reaching $\sim$30~TU (30\% of the mean level) on the East Antarctic plateau. This increasing sensitivity to cosmogenic production at the poles, and especially over Antarctica, is due to injections of stratospheric air masses with greater tritium content and the extremely low water vapour content of this very cold air, combined with the continental and latitudinal effects described above.

\begin{figure}[htb]
\center
\includegraphics[width=\columnwidth]{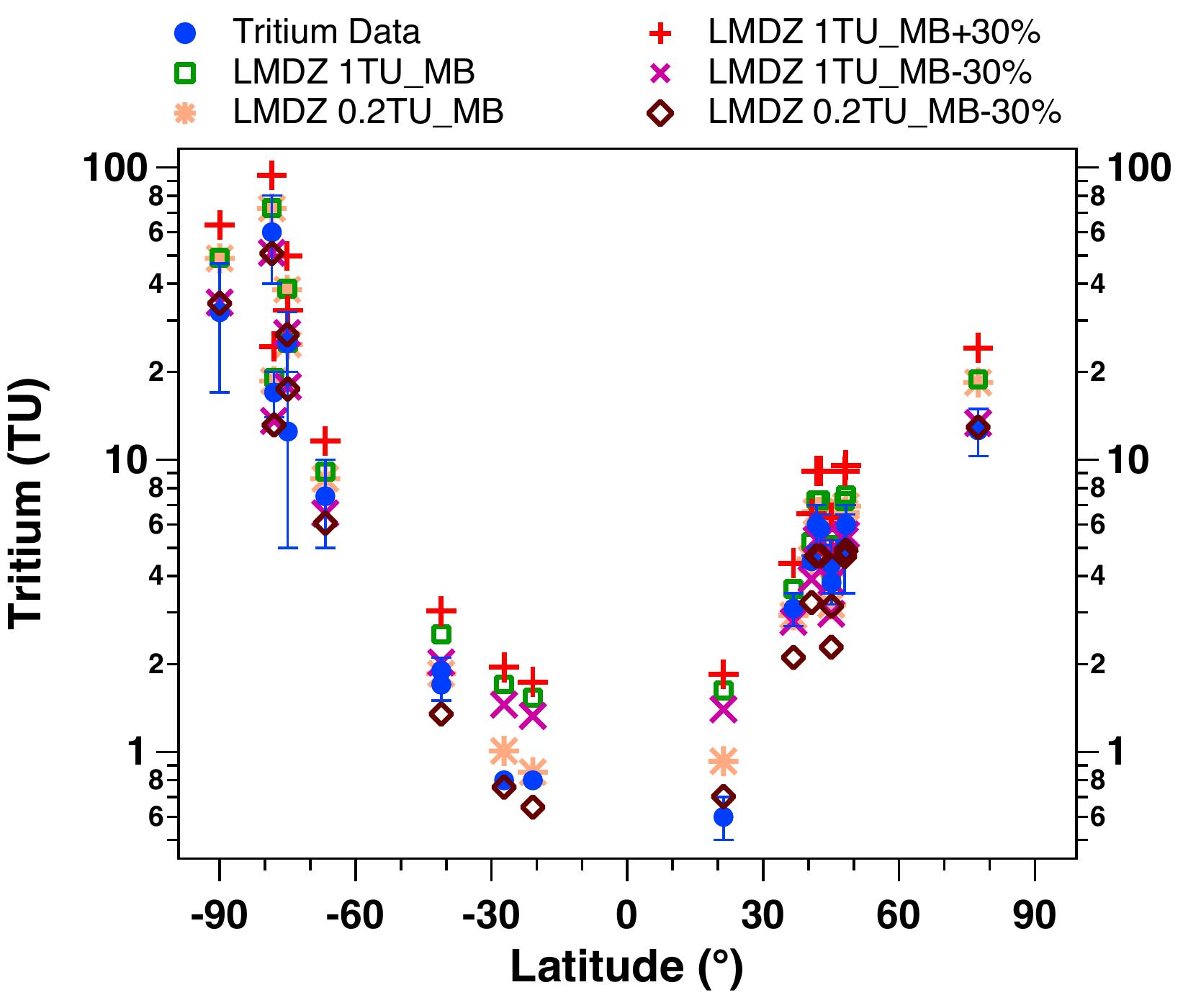} 
\caption{Comparison of the pre-bomb tritium data listed in Table S1 (blue circles) with the tritium in precipitation from the different simulations (coloured symbols) described in Section \ref{simulations}.}
\label{fig7}
\end{figure}

Cosmogenic tritium production and the stratosphere-troposphere flux depend primarily on latitude. To better adjust the production term and the mean level of tritium at the ocean surface, Fig.\ \ref{fig7} shows a latitudinal comparison of the mean tritium in precipitation for the different modelling experiments with the available pre-bomb measurements. The results of the different simulations are computed by performing a bilinear interpolation of the four nearest grid-point values to the sites' coordinates. The latitudinal dependency is clearly visible in Fig.\ \ref{fig7} with a higher tritium content (up to 80~TU) at high latitudes, but very low values at mid-latitudes and in the tropics (a few TU). All simulations are globally in agreement with the data and show the same trend with respect to latitude. However, the tritium concentrations in the ``1TU\_MB'' simulation (green squares) are in the upper range of the data error bars between 60$^{\circ}$S and 90$^{\circ}$S and greater than the measured tritium level between 50$^{\circ}$S and 30$^{\circ}$N. Similarly, the tritium concentrations in the ``1TU\_MB+30\%'' simulation (red + symbols) are too high compared to the data. We find a better agreement with the pre-bomb data by reducing the tritium content at the ocean surface to 0.2~TU (light red stars), especially at low and middle latitudes where the ocean plays a major role in diluting the tritium distribution. For higher latitudes (between 40 and 90$^{\circ}$), the decrease of tritium production by 30\% (magenta $\times$ marks) in the model is the most efficient way to reduce the tritium in precipitation over these areas and to get closer agreement with the measurements. As a reminder, these prescribed values represent both the lower boundaries of the estimated tritium concentration at the ocean surface and MB production rate. This comparison justifies the choice of the ``0.2TU\_MB-30\%'' simulation (brown open diamonds, Section \ref{simulations}) as the reference for LMDZ-iso to give the best reproduction of the annual distribution of natural tritium in the hydrological cycle. 

\subsection{Influence of isotopic fractionation on the continental effect} \label{fractionation}
\begin{figure}[htb]
\center
\includegraphics[width=0.95\columnwidth]{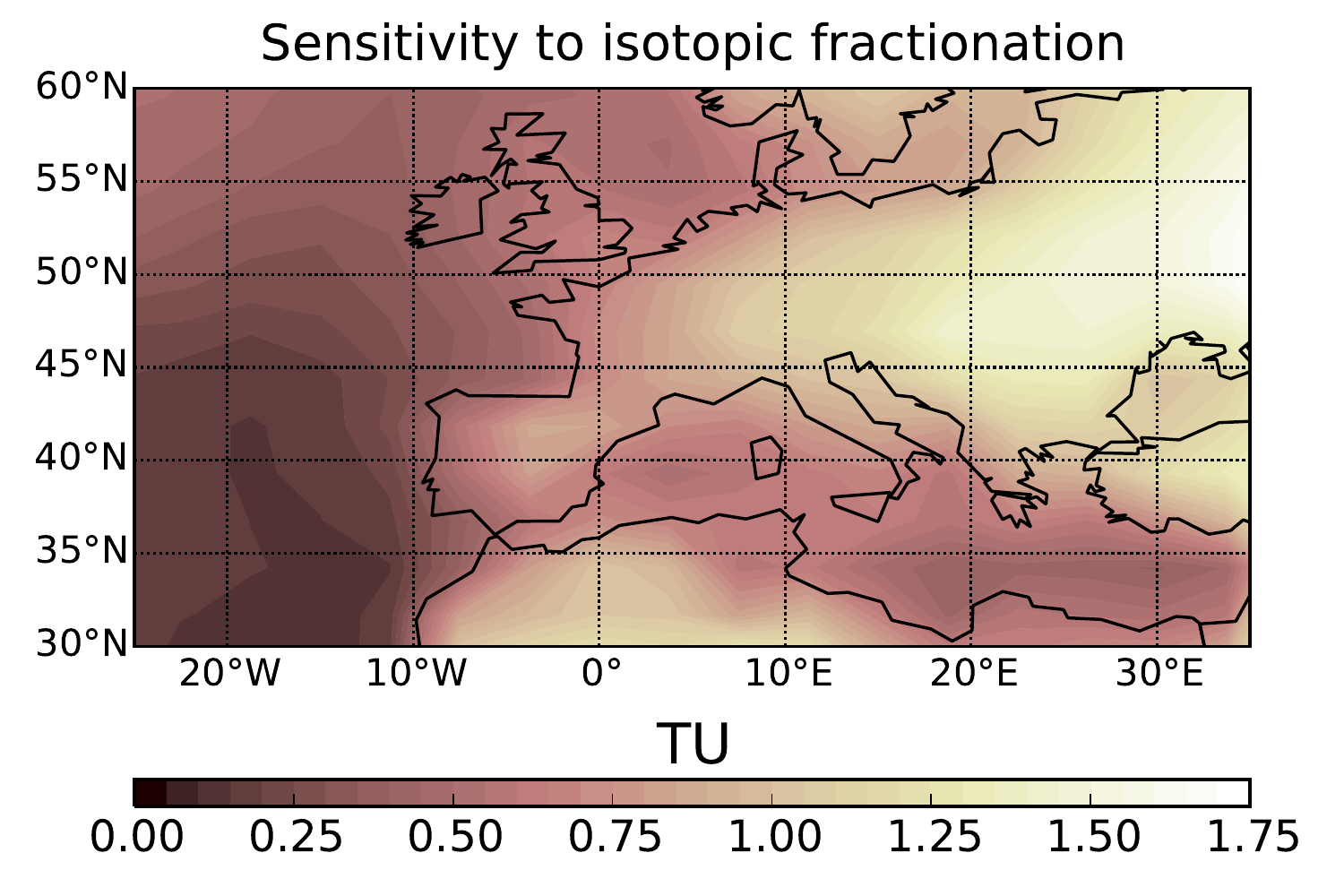} 
\caption{Tritium in precipitation from the ``0.2TU\_MB-30\%'' simulation minus the one from the ``$\alpha$ = 1'' simulation in order to evaluate the sensitivity of the modelled HTO to fractionation processes.}
\label{fig8}
\end{figure}

Cosmogenic production is not the only driver of tritium increase with distance from the coast. The continental effect on tritium concentration is also influenced by isotopic fractionation during water phase changes along water mass trajectories. To decipher the effects of production and water mass distillation on the continental tritium increase, we have performed an experiment without isotopic fractionation. Fig.\ \ref{fig8} displays the difference of tritium in precipitation over Europe between the ``0.2TU\_MB-30\%'' and ``$\alpha$ = 1'' simulations. With fractionation during phase changes, the water in precipitation is globally richer in heavy isotopes, i.e., the HTO/H$_2$O ratio of precipitation is higher. Moreover the difference in tritium level between both simulations increases with the distance from the coast. This increase is due to fractionation during vapor-to-liquid condensation after incorporation of the production of tritium in the troposphere. This means that the continental effect has two origins: tritium production in the high atmosphere and amplification of the upper tropospheric signal by the fractionation from vapour to precipitation. Distillation of water vapour along air mass trajectories acts to slightly reduce the continental gradient in the vapour, but this effect is of minor importance. The effect of fractionation is to increase the continental gradient from the coast (5$^{\circ}$W) to central Europe (15$^{\circ}$E) by 32\%, from 0.43~TU to 1.18~TU.

\subsection{Influence of tritium production and tritium concentration at the ocean surface on the continental effect} \label{europe}
\begin{figure}[htb]
\center
\includegraphics[width=\columnwidth]{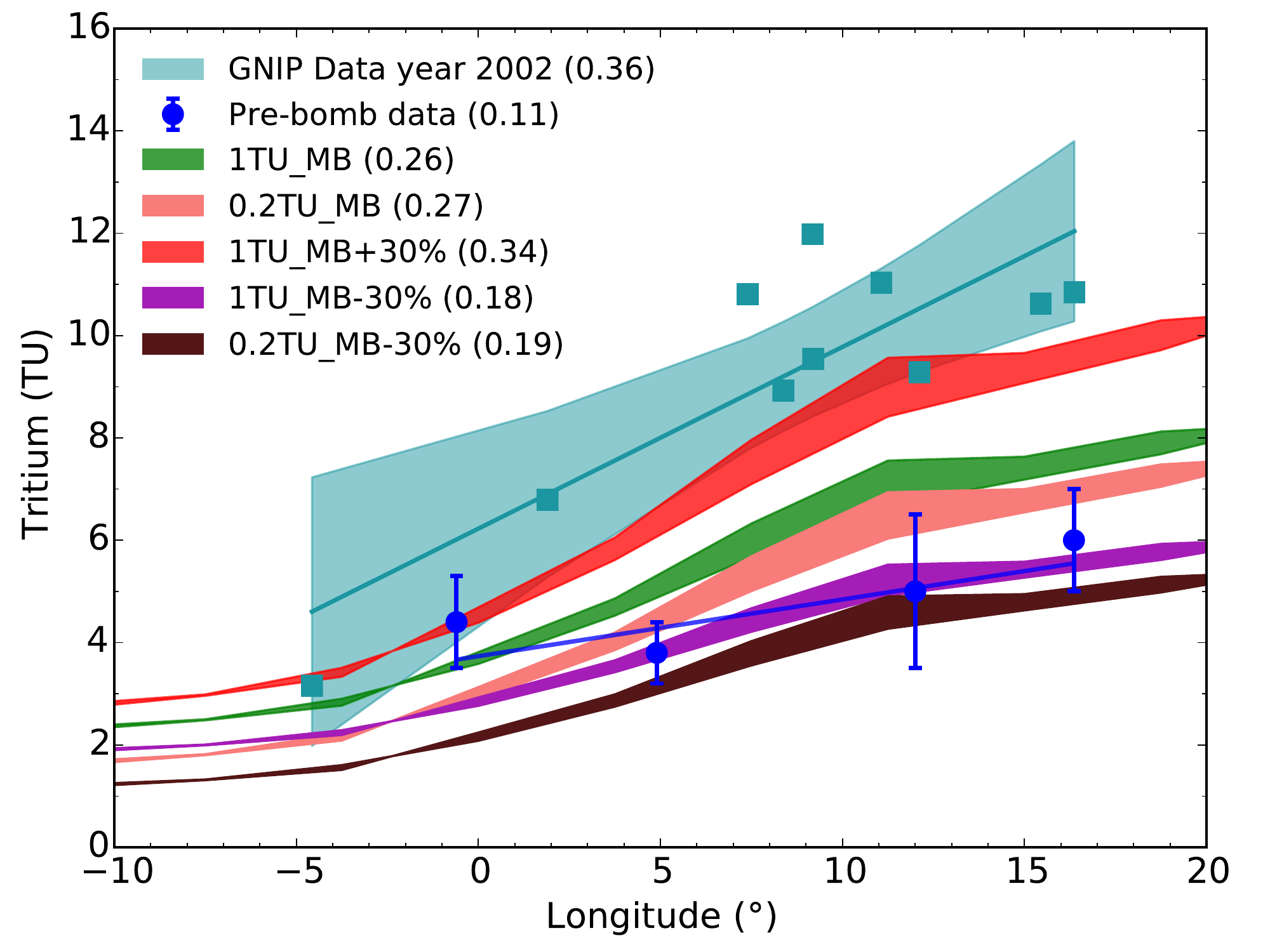} 
\caption{Continental gradient of tritium in precipitation in a latitudinal band between 47 and 49.5$^{\circ}$N for GNIP data from the year 2002 (blue squares plus confidence band at $\pm$1$\sigma$ = 95\%), pre-bomb measurements (blue points), and simulation outputs. The TU -- longitude average slopes (in TU per degree) between 3.75$^{\circ}$W and 15$^{\circ}$E of longitude for each simulation/dataset are given in brackets.}
\label{fig9}
\end{figure}

Fig.\ \ref{fig9} presents a comparison of the tritium levels in precipitation for each simulation/sensitivity test and available dataset between 3.75$^{\circ}$W and 15$^{\circ}$E in a latitudinal band from 47 to 49.5$^{\circ}$N. The continental gradient is greater in the GNIP data compared to that of pure pre-bomb data, showing the residual anthropogenic influence in present day precipitation (average slope coefficients of 0.36 and 0.11~TU per degree for GNIP and pre-bomb data, respectively). Indeed, the influence (even though small) of bomb tritium in the GNIP data increases the concentration of tritium in precipitation especially moving away from the coast where it is progressively diluted by the ocean. 

Comparing the different model runs, we first observe that the ``1TU\_MB +30\%'' simulation has the strongest continental gradient over Europe (red curve: 0.34~TU per degree) and is in much better agreement with the GNIP data than with the pre-bomb measurements. The continental gradients in the ``1TU\_MB'' and ``0.2TU\_MB'' simulations are again too high compared to the pre-bomb observations (green and pink curves: 0.27~TU per degree). In order to reproduce the continental gradient of the pre-bomb data, the production rate would have to be reduced. We obtain a better agreement for modelling experiments ``1TU\_MB-30\%'' and ``0.2TU\_MB-30\%'' with gradients of 0.18 and 0.19~TU per degree respectively, closer to the one deduced from the pre-bomb data (0.11~TU per degree), even if the simulated continental effect always seems slightly too strong. This is especially due to the difference of up to 2~TU in the tritium concentration between our simulations and the pre-bomb measurement in the Gironde Valley (France, [45$^{\circ}$N; 0$^{\circ}$ 36'W], \citet{libby1955}). Note also that we have only four pre-bomb data points to describe the continental effect over Europe, which is probably not enough. Thus, we conclude that the lower range of the tritium production function estimated by \citet{masarik09} should be applied here, i.e., to decrease their calculated values by 30\% (MB-30\%). This procedure will reproduce the continental effect recorded in natural tritium data in Europe reasonably well.

\subsection{Tritium production and the spatial tritium in precipitation distribution in Antarctica} \label{antarctica}
The occurrence of a strong polar vortex in austral winter makes Antarctica the most sensitive region for stratospheric air injections into the troposphere. The use of tritium to study these phenomena is thus of great interest in this region. The tritium in precipitation is quite high in Antarctica, between 20 and almost 100 TU --- confirmed by the measurements in ice or firn. Over this region located at high latitude and at altitude going up to 3500~m, the tritium production rate is higher. Combined with the extreme meteorological conditions, i.e., very cold temperature and very low atmospheric water vapour content (8.25.10$^{-5}$~kg.kg$^{-1}$ above the Antarctic plateau [75$^{\circ}$S; 83$^{\circ}$E] against 0.018~kg.kg$^{-1}$ above the surface ocean [1.27$^{\circ}$N; 180$^{\circ}$W]), the effect of the stratospheric intrusions of air masses with a high tritium content have a maximum effect on the local air tritium concentration, thus explaining this higher level of tritium in snow.

\begin{figure*}[htb]
\center
\includegraphics[width=0.95\textwidth]{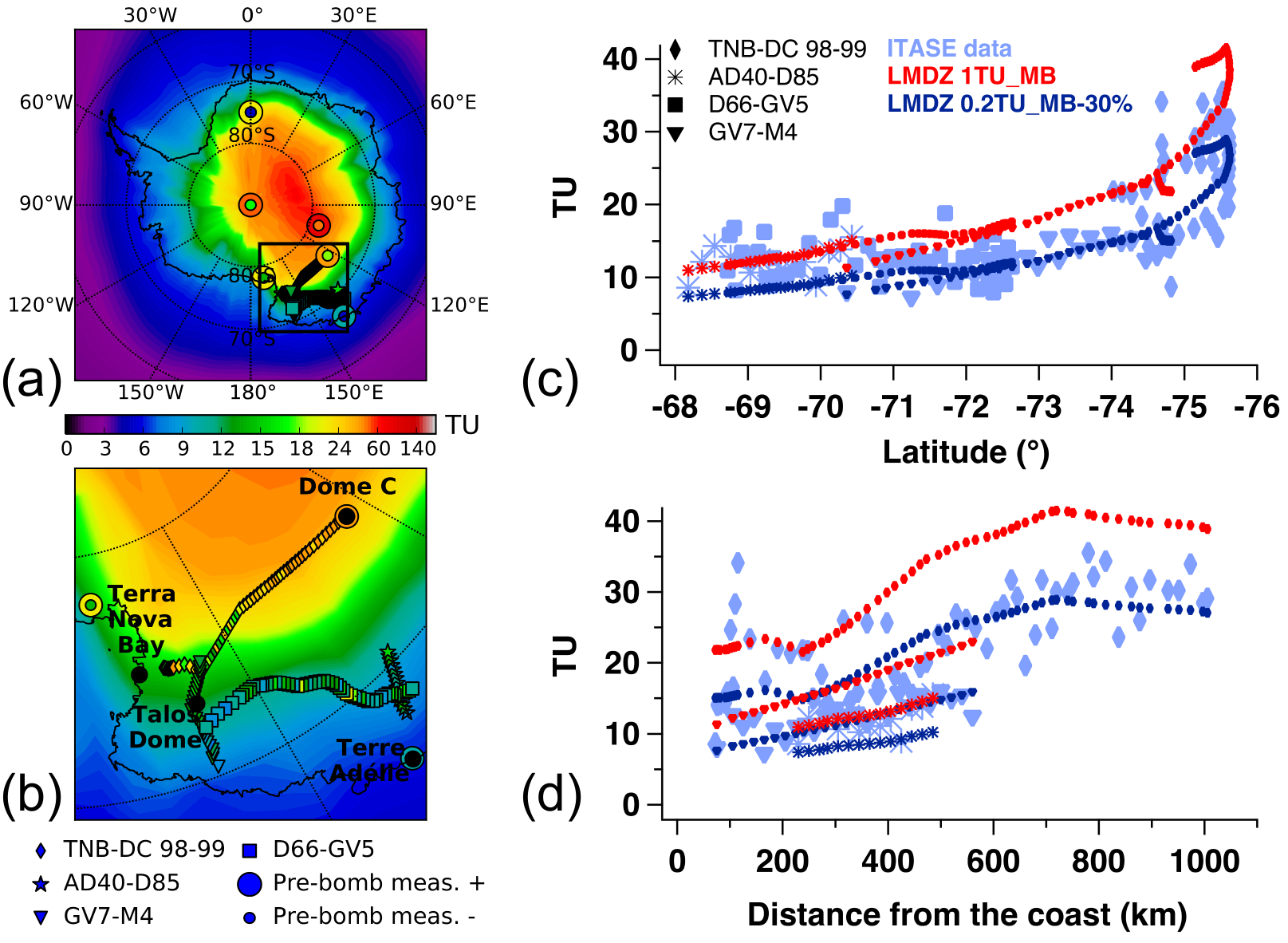} 
\caption[Caption fig 10.]{Comparison of modelled tritium in precipitation with surface snow samples from two ITASE Antarctic traverses \citep{proposito, becagli} and some pre-bomb measurements. (a) Antarctic map from the ``0.2TU\_MB-30\%'' simulation and (b) zooming in the area of interest. (c) Tritium in the ITASE surface snow samples (light blue markers) and modelled tritium in precipitation at these same sites as in the ``1TU\_MB'' simulation (red markers) and the ``0.2TU\_MB-30\%'' simulation (bold-blue markers) as a function of latitude and (d) vs.\ the distance from the coast\footnotemark[1].}
\label{fig10}
\end{figure*}

In agreement with above argument, the ``0.2TU\_MB-30\%'' simulation output (Fig.\ \ref{fig10}a, b) gives the best agreement with snow data. To emphasize their similarities/differences, we have plotted our modelled tritium from the ``1TU\_MB'' and ``0.2TU\_MB-30\%'' simulations and the ITASE data as a function of the latitude and of the distance from the coast (Fig.\ \ref{fig10}c and \ref{fig10}d respectively). As already noticed in the Section \ref{sensitivity}, the evolution of tritium level according to latitude is the same for the different simulations (bold-blue and red markers) and the data (light blue markers). Except for a few anomalous values at the beginning of TNB-DC 98-99 traverse (at 100~km from the coast) already reported by \citet{proposito}, the model also reproduces well the positive trend of tritium in precipitation (tritium content of snow) moving inland and with higher elevation. We observe both in the simulations and in the data tritium level increases until 750~km from the coast, after which a small decrease occurs between 750 and 1000~km. This rise may be due to more efficient winter exchange between stratosphere and troposphere at higher elevations and/or changes in accumulation rate occurring at the different sampling sites. This positive correlation between tritium activity and distance from the coast/elevation has already been observed in Antarctica \citep{merlivat1977, jouzel1979, jouzel1982}.

Unsurprisingly, the ``0.2TU\_MB-30\%'' simulation is in good agreement with the observations made over the Antarctic area. Indeed, especially for the TNB-DC 98-99 and the GV7-M4 traverses, the ``1TU\_MB'' simulation overestimates the tritium concentrations in a range between 4.5 and 14~TU in the 72$^{\circ}$S - 76$^{\circ}$S interval. Moreover, a bias of 10-15~TU with the data from the TNB-DC 98-99 transect is clearly noticeable (red markers in Figs.\ \ref{fig10}c and \ref{fig10}d). It confirms that in Antarctica tritium in precipitation is very sensitive to calculated changes on the tritium production rate, and so to stratospheric injections. Like the GNIP precipitation data, the Antarctic snow samples from these expeditions are also somehow influenced by the residual bomb-tritium. Thus one should expect our simulated tritium level to be lower than those from the data. This lack of discrepancy could indicate that the present-day levels of tritium in Antarctica are not far from pre-bomb values \citep{fourre}. However, as argued in Section \ref{seasonal}, it could be due to the too strong horizontal diffusion of the stratospheric air injections with higher tritium content.

\section{Conclusion and perspectives}\label{conclusion}
We have implemented natural tritium in the LMDZ-iso model. This development is the first GCM modelling of the natural tritium content of water (vapour, precipitation, snow, etc). We have been able to study the spatial and temporal variability of natural tritium in precipitation through an improved physical package and resolution, a long simulation period with proper climatic forcing as well as an updated calculated cosmogenic production rate taken from \citet{masarik09}. In the stratosphere, modelled values of tritium content in water vapour are on the low side of the range of estimations based on the extrapolation of a small number of stratospheric measurements. To see whether or not the model really underestimates the tritium level in the stratospheric water vapour, new stratospheric measurements would be necessary. We have also presented a comprehensive compilation of published data for natural tritium in precipitation in order to evaluate the model performance in terms of spatial and temporal variability. LMDZ-iso reproduces the annual-mean distribution of tritium reasonably well. It captures: (1) the observed tritium enrichment in precipitation moving inland (continental effect), which results from the tritium production term and is further amplified by isotopic fractionation; (2) the observed latitudinal variations, which results from the production rate's latitudinal dependency; and (3) the observed spatial variability of tritium in precipitation over Antarctica in relation to distance from the coast and altitude. The results closest to the pre-bomb observations are found using the ``0.2TU\_MB-30\%'' simulation, i.e., by setting the ocean surface at 0.2~TU and by decreasing the production calculations of \citet{masarik09} by 30\%. This reduction is within the limit of the given uncertainties, suggesting that the average MB values may be overestimated. However, one must keep in mind that by assuming an average value for the solar modulation parameter, the temporal variations of solar modulation were not taken into account.

The seasonal variability, linked to the stratospheric intrusions of air masses with higher tritium content into the troposphere, is correctly reproduced over Antarctica with a clear tritium maximum in austral winter, although with a larger amplitude than in the data. This difference with the data can be explained by stratospheric intrusions that expand too far over the Antarctic continent and the near ocean. LMDZ-iso reproduces the spring maximum of tritium over Europe, but underestimates its value. In addition, the model actually produces a stronger peak in winter that is not observed in the data. The dynamics of stratospheric intrusion into the troposphere and the associated inputs of tritium are still partly an open question: because the tritium production varies with altitude in the stratosphere, the dynamics of the air masses inside the stratosphere may greatly influence tritium levels in tropospheric water vapour. Such effects will be studied in more detail in future work. We will especially focus on the role of detrainment of the condensate by the deep convection depending on its representation in the model and its vertical extension. The modelling of cirrus is also a key parameter, which depends on the microphysics parameterization as the size of ice crystals in the clouds.

In conclusion, the implementation of tritium in a GCM, and subsequent model-data comparison with a large available dataset, is a useful complement to the existing approach of using stable water isotope tracers. The method promises to provide better constraints on: (1) the intrusions and transport of air masses from the stratosphere and (2) the dynamics of the water cycle in the model. The next step will be to implement the bomb-tritium in LMDZ-iso. The atmospheric thermonuclear tests in the 1950s and early 1960s were indeed responsible for the massive injection of tritium in the atmosphere. The detailed information on these nuclear tests (dates, locations, altitudes, yields, etc.) has now been released by governments \citep{unscear} and will be used to reconstruct a realistic atmospheric bomb-tritium input function. Simulations of bomb-tritium will be a very useful tool to evaluate the dynamics of the hydrological cycle in the LMDz-iso model in response to the bomb-tritium transient, and to revisit the issue of the stratospheric injections in relation to the modelled vertical advection. 

\footnotetext[1]{Symbols legends for picture (d) are similar to those in picture (c)}

\section*{Acknowledgements}
We acknowledge two anonymous reviewers for their comments on this manuscript. We thank J. Beer for providing the numerical values of the cosmogenic tritium production. This work was funded by the ERC Grant COMBINISO Project No.\ 306045. LMDZ simulations have been performed on the Ada machine at the IDRIS computing centre under the GENCI project 0292.




\end{document}